\documentclass[prb]{revtex4}
\usepackage{graphicx}

\begin{document}

\title{Upper Pseudogap Phase: Magnetic Characterizations}
\author{Zheng-Cheng Gu and Zheng-Yu Weng}
\affiliation{\textit{Center for Advanced Study, Tsinghua University, Beijing 100084}}
\date{{\small \today}}

\begin{abstract}
It is proposed that the \emph{upper} pseudogap phase (UPP) observed in the
high-$T_{c}$ cuprates correspond to the formation of spin singlet pairing
under the bosonic resonating-valence-bond (RVB) description. We present a
series of evidence in support of such a scenario based on the calculated
magnetic properties including uniform spin susceptibility, spin-lattice
relaxation and spin-echo decay rates, which consistently show that strong
spin correlations start to develop upon entering the UPP, being \emph{%
enhanced} around the momentum ($\pi $, $\pi $) while \emph{suppressed}
around ($0$, $0$). The phase diagram in the parameter space of doping
concentration, temperature, and external magnetic field, is obtained based
on the the bosonic RVB theory. In particular, the competition between the
Zeeman splitting and singlet pairing determines a simple relation between
the \textquotedblleft critical\textquotedblright\ magnetic field, $H_{%
\mathrm{PG}},$ and characteristic temperature scale, $T_{0},$ of the UPP. We
also discuss the magnetic behavior in the \emph{lower} pseudogap phase at a
temperature $T_{v}$ lower than $T_{0}$, which is characterized by the
formation of Cooper pair amplitude where the low-lying spin fluctuations get
suppressed at both ($0$, $0$) and ($\pi $, $\pi $). Properties of the UPP
involving charge channels will be also briefly discussed.
\end{abstract}

\maketitle

\section{Introduction}

The pseudogap phase has been widely regarded as an essential integral part
of the cuprate superconductors with extensive experimental support.\cite%
{timusk} The underlying physics of such a phase has been a central focus in
the study of high-$T_{c}$ problem for many years, and yet no consensus has
been reached concerning its nature due to the very complexity of pseudogap
phenomena observed in magnetic, transport, single-particle, and optical
channels. More and more experimental evidence in recent years further
indicates the existence of two kinds of pseudogap regimes at different
temperatures.\cite{timusk,ps1Bi2212,ps2Bi2212,twops,timusk1,ando}

Among various theoretical proposals for the pseudogap phase, the RVB idea
\cite{anderson} is uniquely interesting, which actually \textquotedblleft
predicted\textquotedblright\ \cite{anderson,kotliar,p.a.li,anderson1} the
existence of a pseudogap state in doped Mott insulators before experiment.
In the RVB picture, neutral spins form the singlet pairs that are condensed
as a spin liquid. The density of states of spin excitations for such a
system generally gets suppressed at low energy, exhibiting a pseudogap
feature as it costs energy to break up the RVB pairs to create spin
excitations. In such a scenario, there is usually no gap in the pure charge
degrees of freedom, and the pseudogap phenomena observed in the charge
transport, angle resolved photoemission spectroscopy (ARPES), and tunnelling
experiments, are all indirectly attributed to the appearance of the \emph{%
spin} gap in the spin degrees of freedom.\cite{p.a.li} For instance, in the
charge transport the strong scattering between the charge carriers and
low-lying spin fluctuations becomes weakened because of the reduction of the
latter in the pseudogap regime; The pseudogap feature exhibited in the ARPES
and tunnelling measurements may be also interpreted as due to the opening of
a pseudogap associated with the spin degrees of freedom.

However, the original RVB description, known as the fermionic RVB (f-RVB)
since it involves the pairing of \emph{fermionic} \textquotedblleft
spinons\textquotedblright\ (neutral $S=1/2$ object),\cite{p.a.li} also
suffers some notable inconsistency with the experiment. Note that the
pseudogap phenomenon has been found in the underdoped regime of the cuprates
\cite{timusk} where the antiferromagnetic (AF) correlations are usually
quite strong. But in an f-RVB description, the AF correlations remain
intrinsically weak, even at half-filling, where the AF long range order
(AFLRO) develops in the cuprates at low temperature. Here the key issue is
not about whether one can construct an AFLRO in the RVB background, which
may be easy to incorporate by a simple mean-field order parameter.\cite%
{hsu,muthu} But the crucial and highly nontrivial issue is whether the \emph{%
whole} low-lying AF spin correlations are intrinsically and sufficiently
strong in an RVB state \cite{wen}, and whether they are capable of
continuously \emph{growing} with reducing temperature or doping as has been
clearly manifested experimentally in, say, the NMR spin-lattice relaxation
rates (see the analysis in Refs. \cite{nmr1,shastry,mmp,ratio1,nmr4}).

Since the pseudogap phase, which involves high-energy/temperature and short
distance physics, may be properly considered as an unstable fixed point
state \cite{anderson1} with intrinsic instabilities towards AFLRO or d-wave
superconductivity at low temperatures, the importance of its correct
description, with regard to the latter, is like that of a Fermi liquid with
regard to the BCS superconducting state. In other words, finding an accurate
and correct description of the pseudogap phase will be rather important for
constructing a sensible low-energy theory for describing the low-temperature
AF and superconducting phases in the cuprates.

\begin{figure}[tbp]
\begin{center}
\includegraphics[width=3in]
{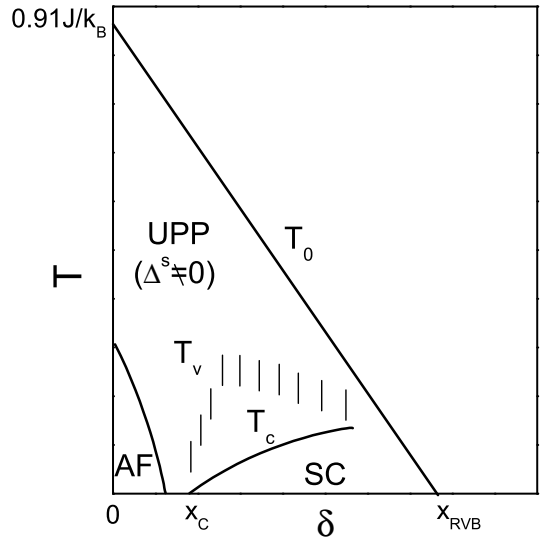}
\end{center}
\caption{The global phase diagram in the b-RVB description (Ref.
\protect\cite{review}). The upper pseudogap phase (UPP) is characterized by
the bosonic RVB pairing order parameter $\Delta ^{s}$ at $T\leq T_{0}$ whose
properties are the main focus of this paper. The antiferromagnetic ordered
phase (AF), the lower pseudogap phase at $T\leq T_{v}$, and the
superconducting phase (SC) at $T\leq T_{c}$ all happen on top of this UPP at
low doping.}
\label{fp}
\end{figure}

In this paper, we show that there does exist a desirable candidate for
characterizing the pseudogap phase based on the RVB picture, in which strong
AF correlations are present as an \emph{intrinsic} and \emph{predominant}
feature. Such an RVB state, known as the bosonic RVB (b-RVB) state,\cite%
{PSMFd} differs from the usual f-RVB states by that it works very well at
half-filling in describing the AF correlations over a wide range of
temperature, including zero temperature where an AFLRO naturally emerges.
Fig. \ref{fp} schematically illustrates the global phase diagram for such a
b-RVB theory,\cite{review} where an \emph{upper }pseudogap phase (UPP) below
the characteristic temperature $T_{0}$ is characterized by the formation of
singlet pairing of neutral spins as denoted by the b-RVB order parameter $%
\Delta ^{s}.$ As illustrated by Fig. 1, the low-temperature phases,
including a \emph{lower} pseudogap phase below $T_{v}$ (also known as the
spontaneous vortex phase \cite{PS1}), a d-wave superconducting phase at $%
T_{c}<T_{v}$ beyond a critical doping concentration $x_{c}$, and an AFLRO
phase near half-filling, can all be regarded as the results of the
low-temperature instabilities from such an UPP.\cite{review}

The main focus of the present work will be the nature of the UPP itself. We
will examine the detailed behavior of uniform spin susceptibility,
spin-lattice relaxation rates, and spin-echo decay rate based on the b-RVB
theory, which will reveal that, as one enters the UPP from above $T_{0}$,
the spin correlations change qualitatively. At the mean-field level,
localized spins are essentially uncorrelated at $T>T_{0}$, resulting in a
Curie-Weiss-like behavior in the uniform spin susceptibility. Below $T_{0}$,
however, finite-range spin correlations start to develop, predominantly
around the AF momentum $\mathbf{Q}_{\mathrm{AF}}=(\pi ,\pi )$, with the
weight being transferred from the momentum $\mathbf{Q}_{0}=(0,0)$. The
latter effect leads to the reduction of uniform susceptibility at $T<T_{0}$.
We show that such magnetic behavior in the UPP is quite consistent with the
experimental measurements in the cuprates.

The above results clearly indicate that the UPP corresponds to a crossover
from a weakly correlated localized spin system at higher temperature into a
strongly \emph{AF}\ correlated spin liquid at lower temperature. This
picture is thus in sharp contrast to the f-RVB picture of the pseudogap,
where the low-energy spin excitations, either around $\mathbf{Q}_{\mathrm{AF}%
}$ or $\mathbf{Q}_{0}$, \emph{all} get suppressed with the opening of the
pseudogap. This latter behavior rather resembles the lower pseudogap phase
of the b-RVB theory at $T\leq T_{v}$ (in Ref. \cite{PSk} it is simply called
the pseudogap phase) than the UPP. But even the distinction between the
pseudogap state of the f-RVB theory and such a lower pseudogap state of the
b-RVB theory is very significant as has been discussed in Ref. \cite{PSk}:
the former is exchange energy driven while the latter is kinetic energy
driven. Here the lower pseudogap phase corresponds to the formation of
Cooper pair amplitude but is short of superconducting phase coherence, which
exists between $T_{v}$ and $T_{c}$ and can be regarded as a vortex liquid
state.\cite{PS1} It was previously pointed out in Ref. \cite{PS1} that such
a spontaneous vortex phase should coincide with the experimentally
discovered Nernst region \cite{nernst} in the high-$T_{c}$ cuprates.

The quantitative phase diagram of the UPP\ is determined in the
three-dimensional parameter space of temperature, doping concentration, and
external magnetic field. The latter introduces the competition between the
Zeeman spin splitting and singlet spin pairing. The mean-field theory will
predict a simple proportional relation between the \textquotedblleft
critical\textquotedblright\ magnetic field $H_{\mathrm{PG}}$, at which the
UPP is destroyed, and $T_{0}$ in zero field. A comparison with experiment
will be made.

The remainder of the paper is arranged as follows. In Sec. II A, the bosonic
RVB description is briefly reviewed. In Sec. II B, the definition of the
upper pseudogap phase is given and its phase diagram is determined. In Sec.
II C and D, its magnetic properties are systematically investigated in the
framework of the b-RVB theory. In Sec. II E, we further briefly discuss the
lower pseudogap phase and related magnetic behavior. Finally, Sec. III is
devoted to conclusion and discussion, where we also briefly discuss the
qualitative behavior of charge channels in the pseudogap phase within the
bosonic RVB description.

\section{Upper pseudogap phase in the bosonic RVB theory}

\subsection{Bosonic RVB description}

The so-called b-RVB state\cite{review} is underpinned by a \emph{bosonic RVB
order parameter} $\Delta _{ij}^{s}$ over a wide range of temperature and
doping as schematically shown in Fig. 1, which defines the UPP. This UPP
(before the emergence of low-temperature AF and superconducting
instabilities) will be the main subject to be examined in the present work.
In the following we shall first discuss its \textquotedblleft
mean-field\textquotedblright\ description based on the $t-J$ model.

In the phase string representation \cite{PS} of the $t-J$ model (see
Appendix A), a natural \textquotedblleft mean-field\textquotedblright\
decoupling of the superexchange term $H_{J}$ is given as follows:\cite{PSMFd}
\begin{equation}
H_{J}\rightarrow H_{s}=-\frac{J}{2}\sum_{\langle ij\rangle \sigma }\Delta
_{ij}^{s}e^{i\sigma A_{ij}^{h}}b_{i\sigma }^{\dagger }b_{j-\sigma }^{\dagger
}+H.c.+\text{ }\mathrm{const.}  \label{hs}
\end{equation}%
where the b-RVB order parameter $\Delta _{ij}^{s}$ is defined in terms of
the bosonic spinon annihilation operator $b_{i\sigma }$ by
\begin{equation}
\Delta _{ij}^{s}=\sum_{\sigma }\left\langle e^{-i\sigma
A_{ij}^{h}}b_{i\sigma }b_{j-\sigma }\right\rangle .  \label{parameter}
\end{equation}%
At half-filling, $\Delta _{ij}^{s}$ is equivalent to the usual
Schwinger-boson mean-field order parameter as $A_{ij}^{h}=0,$ with Eq.(\ref%
{hs}) reducing to the Schwinger-boson mean-field Hamiltonian \cite{half}
which describes the AF correlations fairly well in the regime of $\Delta
_{ij}^{s}\neq 0$ at $T<T_{0}=0.91\mathrm{J/k}_{B}$. Away from half-filling,
a topological link field $A_{ij}^{h}$ emerges in the Hamiltonian (\ref{hs})
as well as in Eq.(\ref{parameter}), which represents the influence of the
nonlocal phase string effect induced by the hole hopping \cite{PS} on the
spin degrees of freedom. It is related to the hole density by the following
gauge invariant relation:\cite{PS}
\begin{equation}
\sum_{<ij>\in C}A_{ij}^{h}=\pi \sum_{l\in \Sigma _{C}}n_{l}^{h}  \label{ah}
\end{equation}%
where $C$ denotes an arbitrary loop on the square lattice that encloses a
region $\Sigma _{C}$, and $n_{l}^{h}$ is the holon number operator at site $%
l $.

The \textquotedblleft mean-field\textquotedblright\ Hamiltonian (\ref{hs})
is by nature a gauge model, which is invariant under the $U(1)$
transformation: $b_{i\sigma }\rightarrow b_{i\sigma }e^{i\sigma \theta _{i}}$
and $A_{ij}^{h}\rightarrow A_{ij}^{h}+(\theta _{i}-\theta _{j})$. It can be
easily shown that the spin rotational symmetry is respected by Eq. (\ref{hs}%
) by verifying $[H_{s}\,,\mathbf{S}]=0$ where the definition of the spin
operator $\mathbf{S}$ in the phase string representation is given in
Appendix A. In the low-temperature regime, where the \emph{bosonic} holons
in the b-RVB theory will experience the Bose condensation such that one may
approximately treat $A_{ij}^{h}$ as describing a uniform flux of the
strength $\delta \pi $ per plaquette ($\delta $ denotes the doping
concentration of holes).\cite{PSMFd} In the high-temperature regime, $%
A_{ij}^{h}$ may be treated as describing randomly distributed static $\pi $
flux tubes of concentration $\delta $ since the holons will behave like
incoherent objects there.\cite{PSk} Different from a usual Jordan-Wigner
phase, the gauge field $A_{ij}^{h}$, which is seen by spinons, is attached
to an \emph{independent} degree of freedom, holons, and therefore the above
approximations are reasonable.

In both limits, the bilinear form of Eq.(\ref{hs}) in terms of the bosonic
spinon operator $b_{i\sigma }$ can be diagonalized by the following
Bogoliubov transformation:\cite{PSMFd}
\begin{equation}
b_{i\sigma }=\sum_{m}[u_{m\sigma }(i)\gamma _{m\sigma }-v_{m\sigma
}(i)\gamma _{m-\sigma }^{\dagger }]  \label{bogoliubov}
\end{equation}%
with
\begin{eqnarray}
u_{m\sigma }(i) &=&u_{m}w_{m\sigma }(i)  \nonumber \\
v_{m\sigma }(i) &=&v_{m}w_{m\sigma }(i)  \label{wavefunction}
\end{eqnarray}%
where $w_{m\sigma }(i)$ satisfies the following eigen equation%
\begin{equation}
\xi _{m}w_{m\sigma }(i)=-\frac{J}{2}\sum_{j=nn(i)}\Delta
_{ij}^{s}e^{-i\sigma A_{ji}^{h}}w_{m\sigma }(j)  \label{eig}
\end{equation}%
in which $j=nn(i)$ denotes the four nearest neighbors (nn) of site $i.$ One
has $u_{m}=\frac{1}{\sqrt{2}}\sqrt{\frac{\lambda }{E_{m}}+1}$ and $v_{m}=%
\mathrm{sgn}(\xi _{m})\frac{1}{\sqrt{2}}\sqrt{\frac{\lambda }{E_{m}}-1},$
where $E_{m}=\sqrt{\lambda ^{2}-\xi _{m}^{2}}$ is the spinon spectrum. The
Lagrangian multiplier $\lambda $ is determined by enforcing the average
constraint $\left\langle \sum\nolimits_{\sigma }b_{i\sigma }^{\dagger
}b_{i\sigma }\right\rangle =1-\delta ,$ which leads to
\begin{eqnarray}
2-\delta &=&\frac{1}{N}\sum_{m}\frac{\lambda }{E_{m}}\coth \frac{\beta E_{m}%
}{2}  \label{lambda0} \\
\sum_{\langle ij\rangle }|\Delta _{ij}^{s}|^{2} &=&\sum_{m}\frac{\xi _{m}^{2}%
}{JE_{m}}\coth \frac{\beta E_{m}}{2}  \label{delta}
\end{eqnarray}%
where $\beta =1/k_{B}T$ and the last equation is obtained by the
self-consistent condition (\ref{parameter}) for the RVB order parameter.
Note that a Bose condensed term $n_{BC}^{b}$ related to the AFLRO \cite%
{PSMFd} at $T=0$ and half-filling has been dropped in Eq.(\ref{lambda})
since we shall mainly be interested in the high-temperature behavior.

All the nontrivial effect of doping is reflected in the eigen equation (\ref%
{eig}) where the phase string effect induced by hopping enters via the
topological gauge field $A_{ij}^{h}$. Note that the spinon wave function $%
w_{m\sigma }(i)$ should vanish at the hole sites (where $\Delta _{ij}^{s}=0$%
) due to the no double occupancy condition. Previously,\cite{PSMFd} such an
equation has been solved by a simple mean-field choice $\Delta _{ij}^{s}$ [$%
j=nn(i)$]$=\Delta ^{s}$ with the relaxed constraint condition such that the
spinons can go any sites.\ In the following we still relax the no double
occupancy constraint on average in Eq.(\ref{eig}), but with a replacement of

\begin{equation}
J\rightarrow J_{\text{eff}}=(1-2g\delta )J  \label{jeff}
\end{equation}%
to represent the average effect of the reduction of the superexchange
coupling around holes: if holes are static, each of which will simply break
two $nn$ links in each direction such that $g=1$ in the dilute hole limit
(when the average hole-hole distance is much larger than the $nn$ links).
Generally $g>1$ for a moving hole since the suppression of $\Delta _{ij}^{s}$%
\ around each hole extends more than the four broken $nn$ bonds ($e.g.,$
from the singular twist by $A_{ij}^{h}$ around each hole). \ One may thus
approximately rewrite Eq.(\ref{eig}) as
\begin{equation}
\xi _{m}w_{m\sigma }(i)=-J_{s}\sum_{j=nn(i)}e^{-i\sigma
A_{ji}^{h}}w_{m\sigma }(j)  \label{eig1}
\end{equation}%
in which
\[
J_{s}\equiv \frac{J_{\text{eff}}\Delta ^{s}}{2}
\]%
and Eq.(\ref{delta}) can be consistently rewritten as
\begin{equation}
\Delta ^{s}=\frac{1}{4N}\sum_{m}\frac{\xi _{m}^{2}}{J_{s}E_{m}}\coth \frac{%
\beta E_{m}}{2}.  \label{delta1}
\end{equation}%
Note that the same mean-field equations have been obtained in Ref. \cite%
{PSMFd} with a slightly different definition, \emph{i.e.,} with $\Delta ^{s}$
replaced by $\Delta ^{s}/(1-2g\delta )=\Delta _{1}^{s}$ (with $g=1)$ in Ref.
\cite{PSMFd}.

\subsection{Upper pseudogap phase}

The UPP is defined by the formation of the b-RVB pairing with $\Delta
^{s}\neq 0.$ Its high-temperature boundary at $\Delta ^{s}=0$ is depicted by
a characteristic temperature $T_{0}$ as illustrated in Fig. \ref{fp}. In the
following, we determine it based on the mean-field theory outlined above.

\ According to Eq.(\ref{lambda0}), one finds
\begin{equation}
2-\delta =\coth \frac{\lambda }{2k_{B}T_{0}}  \label{upp1}
\end{equation}%
by noting that $\xi _{m}\rightarrow 0$ at $\Delta ^{s}\rightarrow 0.$
Consequently, Eq.(\ref{delta1}) reduces to $1=(2-\delta )\frac{1}{2N}%
\sum_{m}(\xi _{m}/\Delta ^{s})^{2}/J_{\text{eff}}\lambda $, which gives rise
to
\begin{equation}
\lambda =\frac{2-\delta }{2}J_{\text{eff}}  \label{lambda}
\end{equation}%
by further identifying
\begin{equation}
\frac{1}{N}\sum_{m}(\xi _{m}/\Delta ^{s})^{2}=J_{\text{eff}}^{2}
\label{eig2}
\end{equation}%
in terms of Eq.(\ref{eig1}). Finally one obtains%
\begin{equation}
k_{B}T_{0}=\left( \frac{1-\frac{\delta }{2}}{\ln \frac{3-\delta }{1-\delta }}%
\right) J_{\mathrm{eff}}\text{ \ .}  \label{T0}
\end{equation}%
It is interesting to note that the gauge field $A_{ij}^{h}$ does not
explicitly appear in $T_{0}$ because of the general relation (\ref{eig2}),
indicating that the phase string effect, which is crucial to the low
temperature (low energy) physics, actually plays no role in determining the
phase boundary of the UPP. Of course, the hopping effect still enters in $%
T_{0}$ via the renormalized factor $g$ in $J_{\mathrm{eff}}$ which shall be
the only adjustable parameter depending on the detailed physics of local
hopping. We find that the phase diagram and magnetic properties to be
studied below are actually not sensitive to $g$ except for the
characteristic concentration $x_{\mathrm{RVB}}$ at which $J_{\mathrm{eff}}$
is extrapolated to zero at $T=0$.

\begin{figure}[tbp]
\begin{center}
\includegraphics[width=3in]
{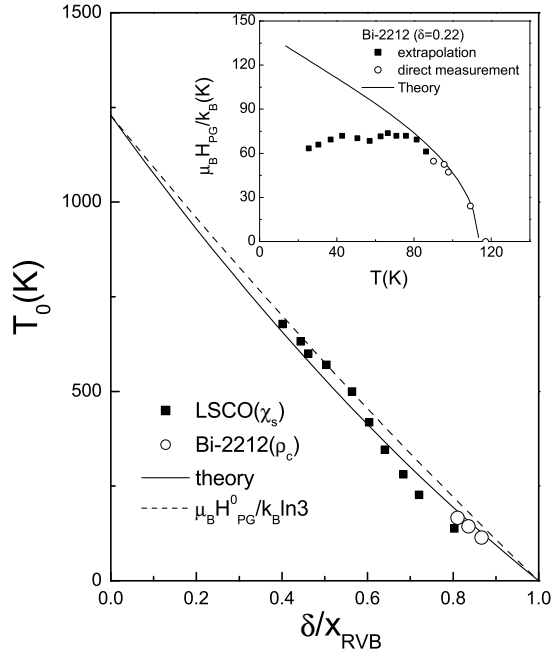}
\end{center}
\caption{The characteristic temperature $T_{0}$ of the UPP versus $\protect%
\delta /x_{\mathrm{RVB}}$. Solid line: the present theory; Full squares:
determined from the uniform spin susceptibility $\protect\chi _{s}$ in LSCO
compound;\protect\cite{LSCO} Open circles: determined from the c-axis
magneto-resistivity ($\protect\rho _{c})$ measurement in Bi-2212 compound;%
\protect\cite{pszeeman} The dashed line shows the scaling relation of the
zero-temperature critical field $H_{\mathrm{PG}}^{0}$ with $T_{0}$ as
predicted by the theory. Inset: the critical field $H_{\mathrm{PG}}$ as a
function of temperature at $\protect\delta =0.22$. The experiment data from
the c-axis transport in Bi-2212 (Ref. \protect\cite{pszeeman}) are also
shown by the open and full squares. }
\label{fT0}
\end{figure}

Fig. \ref{fT0} shows $T_{0}$ (solid curve) as a function of $\delta /x_{%
\mathrm{RVB}}$ with $J=1350$ \textrm{K.} The experimental data determined by
the uniform spin susceptibility measurement in $\mathrm{LSCO}$ \cite%
{scaling,LSCO} (see the discussion in the next section) are shown by the
full squares, which are in good agreement with the theoretical curve.
Furthermore, the open circles are independently determined from the c-axis
transport \cite{pszeeman} in the overdoped regime (see discussion below).
Note that the theoretical curve $T_{0}$ versus $\delta /x_{\mathrm{RVB}}$ in
Fig. 2 is not sensitive to the choice of $g.$ Furthermore, one can use the
above experimental data \cite{scaling,LSCO,pszeeman} to fix $x_{\mathrm{RVB}%
} $ at $x_{\mathrm{RVB}}=1/2g=0.25$ ($g=2$). We shall then choose $x_{%
\mathrm{RVB}}=0.25$ throughout the rest of paper without any more adjustable
parameter.

For the b-RVB origin of the UPP, the Zeeman splitting due to the external
magnetic field can effectively destroy the singlet pairing of spins in the
strong field limit. Since the orbit part of the neutral spins does not
couple to the external field directly, the Zeeman splitting will be the only
direct field effect on the RVB background. It thus provides a direct probe
of the RVB nature of the UPP in, say, the overdoped regime, where the
critical field strength may be within the experimental accessible range. In
the following we consider the Zeeman effect in the UPP.

Apply an external magnetic field $\mathbf{H}$ along a spin z-axis (which is
not necessarily perpendicular to the lattice plane). A spin Zeeman energy
term should be then added to $H_{s}$ in Eq.(\ref{hs}):
\begin{equation}
-2\mu _{B}\sum_{i}S_{i}^{z}H=-\mu _{B}H\sum_{\sigma }\sigma \gamma _{m\sigma
}^{\dagger }\gamma _{m\sigma }
\end{equation}%
Consequently the spinon excitation spectrum is modified by
\begin{equation}
E_{m}^{\sigma }=E_{m}-\sigma \mu _{B}H
\end{equation}%
which now explicitly depends on the spin index $\sigma $. Then the
mean-field equation (\ref{upp1}) at $\Delta ^{s}\rightarrow 0$ is modified
to:
\begin{equation}
2-\delta =\frac{1}{2}\sum_{\sigma }\coth \frac{E^{\sigma }}{2Tk_{B}}
\label{upp3}
\end{equation}%
with $E^{\sigma }\equiv \lambda -\sigma \mu _{B}H$, while Eq.(\ref{lambda})
remains the same. From these equations, we can easily obtain the following
relation between $T_{0}$ at zero field and the zero-temperature
\textquotedblleft critical\textquotedblright\ field $H_{PG}^{0}$ at which $%
\Delta ^{s}$ vanishes:%
\begin{equation}
\mu _{B}H_{\mathrm{PG}}^{0}=\ln \left( \frac{3-\delta }{1-\delta }\right)
k_{B}T_{0}\text{ \ }  \label{h0}
\end{equation}%
with the coefficient only weakly dependent on the doping concentration. In
Fig. \ref{fT0}, $\mu _{B}H_{PG}^{0}/k_{B}\ln 3$ is plotted as the dashed
curve which scales with the zero-field $T_{0}$ fairly well, which predicts%
\begin{equation}
\mu _{B}H_{\mathrm{PG}}^{0}\simeq 1.1k_{B}T_{0}\text{ \ .}  \label{h01}
\end{equation}

In general, the temperature dependence of the \textquotedblleft
critical\textquotedblright\ field $H_{\mathrm{PG}}(T)$ can be obtained based
on Eqs. (\ref{upp3}) and (\ref{lambda}). In the inset of Fig. \ref{fT0}, $H_{%
\mathrm{PG}}$ versus $T$ at $\delta =0.22$ is shown together with the
experimental data obtained from the c-axis magneto transport measurements.%
\cite{pszeeman} We see that the \emph{high-temperature} experimental data
(open circles) fit the theoretical curve very well without any additional
adjustable parameter. Furthermore the zero-field $T_{0}$ determined by the
\emph{same }experiments is also in good agreement with the theory as shown
(open circles) in the main panel of Fig. \ref{fT0} as mentioned above. One
may also note that the experimental $H_{\mathrm{PG}}(T)$ starts to deviate
from the theoretical curve in the inset (full squares) as the temperature is
further lowered and saturated to approximately the half of the predicted
number (which implies $\mu _{B}H_{\mathrm{PG}}^{0}\simeq k_{B}T_{0}/2)$.
However, we would like to point out that such a deviation occurs only for
those data (full squares) which have been obtained by \emph{extrapolation}
in the experimental measurement \cite{pszeeman} and therefore may not be as
reliable as the higher temperature ones (open squares) in the inset of Fig. %
\ref{fT0}. We believe that further experiments be needed in order to
convincingly verify (falsify) the present prediction (\ref{h01}).

\begin{figure}[tbp]
\begin{center}
\includegraphics[width=3.5in]
{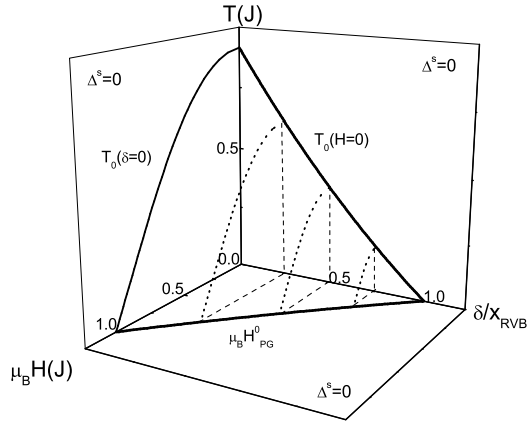}
\end{center}
\caption{The phase diagram of the UPP in the parameter space of doping,
temperature, and external magnetic field, as calculated based on the
mean-field bosonic RVB theory. }
\label{f3d}
\end{figure}

Finally the boundaries of the UPP in the three dimensional parameter space
of doping concentration, temperature, and magnetic field determined based on
the b-RVB mean-field theory are shown in Fig. \ref{f3d}.

\subsection{Uniform spin susceptibility}

The RVB nature of the UPP is clearly manifested in the magnetic properties.
We first consider uniform spin susceptibility $\chi _{s}$ in the following,
which can be easily derived based on the above mean-field description in the
presence of magnetic field. In terms of the total spin magnetic moment
\begin{equation}
M=\mu _{B}\sum_{m}[n(E_{m}^{+})-n(E_{m}^{-})]\text{ \ \ \ \ }  \label{sus3}
\end{equation}%
[where $n(\omega )=1/(e^{\beta \omega }-1)$ is the Bose distribution
function], the uniform spin susceptibility $\chi _{s}$ per cite as defined
by $\chi _{s}=\frac{M}{NH}\mid _{H\rightarrow 0}$ is found by
\begin{equation}
\chi _{s}=\frac{2\mu _{B}^{2}\beta }{N}\sum_{m}n(E_{m})[1+n(E_{m})]\text{ \
\ \ \ .}  \label{sus4}
\end{equation}

\begin{figure}[tbp]
\begin{center}
\includegraphics[width=3.5in]
{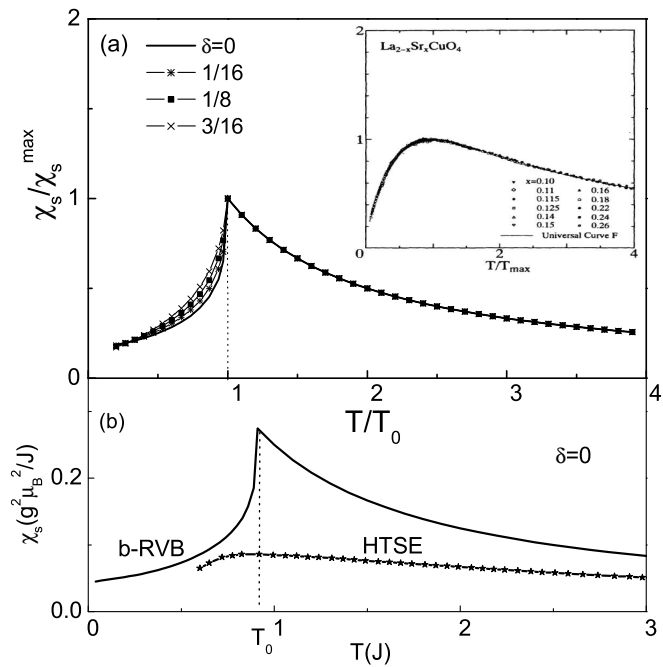}
\end{center}
\caption{(a) The calculated uniform spin susceptibility $\protect\chi _{s}$
scaled with the maximum $\protect\chi _{s}^{\mathrm{\max }}$ at $T_{0}$
versus $T/T_{0}$, which follows an approximately doping-dependent curve.
Inset: The experimental data in Ref. \protect\cite{scaling,LSCO} which
collapse into a universal scaling curve plotted in the same fashion as in
the main panel. (b) The theoretical $\protect\chi _{s}$ at half-filling
(solid) and the one obtained by the high temperature series expansion
(HTSE). The latter fits the experimental scaling curve in the inset of (a)
very well.\protect\cite{scaling,LSCO} }
\label{fsus}
\end{figure}

Since we are mainly interested in the high-temperature regime well above the
superconducting phase, the holes are incoherent objects such that $%
A_{ij}^{h} $ can be approximately treated as describing randomly distributed
$\pi $ flux tubes of concentration $\delta $ as discussed before. Then we
can numerically calculate $\chi _{s}$ based on the mean-field equations
given in Sec. II A, which is averaged under different random configurations
of $A_{ij}^{h}.$ A similar computation has been done before to explore the
crossover from the lower to upper pseudogap phases,\cite{PSk} but not in the
region up close to $T_{0}$. (In the following calculations, the largest
sample size is $32\times 32$ lattice and the sample size is not very
important as mainly the high temperature properties are concerned.)

The calculated $\chi _{s}$ is presented in the main panel of Fig. \ref{fsus}
(a) at different doping concentrations. Note that $\chi _{s}$ reaches a
maximum value $\chi _{s}^{max}$ at temperature $T_{0}$ where the RVB order
parameter $\Delta ^{s}$ vanishes. At $T>T_{0}$, $\chi _{s}$ follows a Curie-$%
1/T$ behavior as spins become free moments at the mean-field level. The
curves in Fig. \ref{fsus} (a) are presented as $\chi _{s}/\chi _{s}^{\max }$
versus $T/T_{0}$, which approximately collapse onto a single curve
independent of doping. For comparison, the inset shows the experimental data
obtained in \textrm{LSCO }compounds which are plotted in the same way as in
the main panel with a very good collapsing onto a universal scaling curve;%
\cite{scaling,LSCO} And the peak positions decides the experimental
pseudogap temperature $T_{0}$'s at different dopings, which are presented in
Fig. \ref{fT0}.

In Fig. \ref{fsus} (b), the calculated $\chi _{s}$ versus $T$ at $\delta =0$
is shown together with the high temperature series expansion (HTSE) result%
\cite{HTSE} (note that here the calculated $\chi _{s}$ is rescaled by a $2/3$
numerical factor as used in the Schwinger-boson approach to restore the sum
rule\cite{half}). It is noted that the experimental scaling curve actually
coincides with the half-filling HTSE very well.\cite{scaling,LSCO} Thus one
can clearly see the overall qualitative agreement between the bosonic RVB
theory and the experiment from Figs. \ref{fsus} (a) and (b). Note that the
mean-field $\chi _{s}$ deviates from the HTSE result prominently around $%
T_{0}$ where the latter is a much smoother function of $T.$ It reflects the
fact that $T_{0}$ is only a crossover temperature and the vanishing $\Delta
^{s}$ does not represent a true phase transition. Obviously, the amplitude
fluctuations beyond the mean-field $\Delta ^{s}$ have to be considered in
order to better describe $\chi _{s}$ in this regime. $T_{0}$ determined in
the mean-field theory is quite close to the HTSE result, indicating the
crossover temperature itself can still be reasonably decided by the
mean-field bosonic RVB description given above. The comparison of $T_{0}$
between the theory and experiment has been already presented in Fig. \ref%
{fT0} and discussed in the previous section.

\subsection{Spin-lattice relaxation rate and spin-echo decay rate}

The NMR spin-lattice relaxation rate of nuclear spins is determined by\cite%
{moriya}
\begin{equation}
\frac{1}{T_{1}}=\frac{2k_{B}T}{g^{2}\mu _{B}^{2}N}\sum_{\mathbf{q}}F(\mathbf{%
q})^{2}\left. \frac{\chi _{zz}^{\prime \prime }(\mathbf{q},\omega )}{\omega }%
\right\vert _{\omega \rightarrow 0^{+}}  \label{NMRG1}
\end{equation}%
where the form factor $F(\mathbf{q})^{2}$ comes from the hyperfine coupling
between nuclear spin and spin fluctuations. For example, for planar $^{63}%
\mathrm{Cu(2)}$ nuclear spins in the cuprates, with the applied field
perpendicular to the $\mathrm{CuO}_{2}$ plane, the form factor $F(\mathbf{q}%
)^{2}$ is found to be:\cite{shastry,mmp}
\begin{equation}
^{63}F(\mathbf{q})^{2}=[A_{\perp }+2B(\cos q_{x}a+\cos q_{y}a)]^{2}
\label{factorCu1}
\end{equation}%
where the hyperfine couplings $A_{\perp }$ and $B$ are estimated as $%
A_{\perp }/B\simeq 0.84,$ $B\simeq 3.8\times 10^{-4}\mathrm{meV}$ \cite{mmp}
[These coefficients may slightly vary among $\mathrm{YBCO}$ and $\mathrm{LSCO%
}$ compounds]. For planar $^{17}\mathrm{O(2)}$ nuclear spins, one has
\begin{equation}
^{17}F(\mathbf{q})^{2}=2C^{2}[1+\frac{1}{2}(\cos q_{x}a+\cos q_{y}a)]
\label{factorO1}
\end{equation}%
with $C\simeq 0.87B$. Due to the fact that $^{17}F(\mathbf{q})^{2}$ vanishes
at the AF wave vector $\mathbf{Q}_{\mathrm{AF}}=(\pi ,\pi )$, while $^{63}F(%
\mathbf{q})^{2}$ is peaked at $\mathbf{Q}_{\mathrm{AF}}$, a combined
measurement of $1/^{63}T_{1}$ and $1/^{17}T_{1}$ will thus provide important
information about the AF correlations at low frequency $\omega \rightarrow 0$%
.

\begin{figure}[tbp]
\begin{center}
\includegraphics[width=3.5in]
{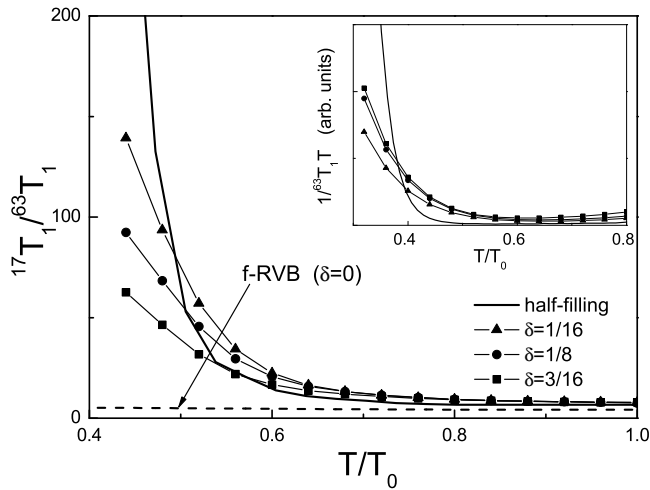}
\end{center}
\caption{$^{17}T_{1}/^{63}T_{1}$ vs. temperature at different doping
concentrations in the upper pseudogap phase of the b-RVB state. The dashed
line shows the result of an f-RVB state ($\protect\pi $ flux phase) at
half-filling. The inset shows the non-Korringa behavior of $1/^{63}T_{1}T$
in the b-RVB state at various dopings (the symbols are the same as in the
main panel).}
\label{fnmr}
\end{figure}

Based on the bosonic RVB mean-field equations outlined in Sec. II A, the
spin-lattice relaxation rates, $1/^{63}T_{1}$ and $1/^{17}T_{1},$ for the
planar copper and oxygen nuclear spins, can be straightforwardly computed as
shown in Appendix B. The results are presented in Fig. \ref{fnmr}. It shows
that the ratio $^{17}T_{1}/^{63}T_{1}$, which is a constant above $T_{0}$,
starts to increase with reducing temperature below $T_{0}$. At lower
temperature, $T/T_{0}<0.5$, such a ratio arises sharply. For example, $%
^{17}T_{1}/^{63}T_{1}$ diverges at $\delta =0$ as a true AFLRO exists at $%
T=0;$ And it can still reaches about $100$ in the low temperature limit at $%
\delta =0.125$, all qualitatively consistent with the experimental
observation in the cuprates.\cite{ratio1} As pointed out above, such
behavior clearly demonstrates that strong low-lying AF correlations around $%
\mathbf{Q}_{\mathrm{AF}}$ develop in the UPP, leading to the simultaneous
enhancement of $1/^{63}T_{1}$ and the cancellation in $1/^{17}T_{1}$. In the
inset of Fig. \ref{fnmr}, $1/^{63}T_{1}T$ has been plotted, which is also
qualitatively consistent with the experiment, \cite{mmp,ratio1,nmr4} but
deviates from the conventional Korringa behavior $1/^{63}T_{1}T\sim const$
for a Fermi liquid system, thanks to the strong AF fluctuations of spins in
the UPP below $T_{0}$.\cite{remark1} By contrast, the ratio $%
^{17}T_{1}/^{63}T_{1}$ in an f-RVB mean-field state (the $\pi $ flux phase)
at half-filling remains flat over the whole temperature region as shown by
the dashed line in Fig. \ref{fnmr}, indicating the absence of any
significant AF correlations around $\mathbf{Q}_{\mathrm{AF}}$ in its
pseudogap regime. In combination with the reduced uniform spin
susceptibility,\cite{PSk} one sees that in the pseudogap of the f-RVB, the
low-energy spin excitations, either around $\mathbf{Q}_{\mathrm{AF}}$ or $%
\mathbf{Q}_{0}=(0,0)$, all get suppressed with the opening of the pseudogap.

The above results clearly show that the UPP in the bosonic RVB state
corresponds to the crossover from a weakly correlated localized spin
assembly at higher temperature into a strongly \emph{AF}\ correlated spin
liquid at lower temperature. The peculiar feature of the bosonic RVB
description is that although the formation of bosonic RVB singlet pairing
suppresses the spin correlations at $\mathbf{Q}_{0}$ below $T_{0}$, it also
leads to the \emph{enhancement} of the low-energy spin correlations near AF
momentum $\mathbf{Q}_{\mathrm{AF}}$. Such a feature of the UPP\ is
significantly different from the pseudogap phase in the f-RVB mean-field
description, but is strongly supported by the NMR measurements.\cite%
{nmr1,shastry,mmp,ratio1,nmr4}

Finally, we further examine the spin-echo decay rate $1/T_{2G}$, which is
related to the static AF correlations via the real part of spin
susceptibility function by:\cite{T2G1}
\begin{equation}
\left(\frac{1}{T_{2G}}\right)^2=\frac{0.69}{8\hbar^2}\frac{1} {%
(\hbar\gamma_e)^4}\left\{\frac{1}{N}\sum_\textbf{q}F_{eff}(\mathbf{q}%
)^4\chi^\prime_{zz}(\mathbf{q})^2 -\left[\frac{1}{N}\sum_\textbf{q}F_{eff}(%
\mathbf{q})^2\chi^\prime_{zz}(\mathbf{q})\right]^2\right\}  \label{T2}
\end{equation}
where the factor $F_{\mathrm{eff}}(\mathbf{q})$ is
\begin{equation}
F_{\mathrm{eff}}(\mathbf{q})=A_{\parallel }+2B(\cos q_{x}a+\cos q_{y}a)
\label{factorT2}
\end{equation}%
with $A_{\parallel }\simeq -4B$, such that $F_{\mathrm{eff}}(\mathbf{q})$ is
peaked at $\mathbf{Q}_{\mathrm{AF}}$ and vanish at $\mathbf{Q}_{0}$.

\begin{figure}[tbp]
\begin{center}
\includegraphics[width=3.5in]
{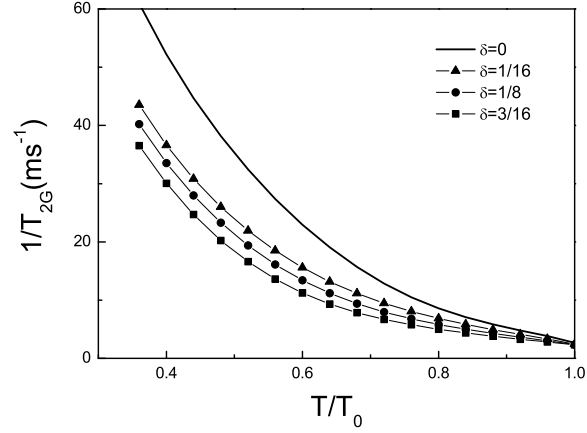}
\end{center}
\caption{$1/T_{2G}$ vs. temperature in the upper pseudogap phase below $%
T_{0} $ at different doping concentrations.}
\label{3}
\end{figure}
Similar to $1/T_{1}$, the detailed expression of $1/T_{2G}$ in the bosonic
RVB theory is given in Appendix C. In Fig.\ref{3}, the calculated $1/T_{2G}$
at different doping concentrations show that $1/T_{2G}$ begins to increase
with reducing temperature below $T_{0}$. Such behavior has been also
observed in the experiment,\cite{T2G2,T2G3,T2G4} which once again clearly
illustrates the picture that the strong AF correlations start to develop in
the UPP.

\subsection{Lower Pseudogap Phase}

So far we have been focused on the UPP, which is the high-temperature phase
in the bosonic RVB description. As stressed in the Introduction, there can
be several different low-temperature phases growing out of this background
(Fig. 1).\cite{review} One particular phase we wish to discuss below is the
so-called spontaneous vortex phase \cite{PS1} which can be properly
classified as the lower pseudogap phase in the present approach. In this
phase, the holon condensation occurs and Cooper pair amplitude forms, but
the system is still short of superconducting phase coherence which can be
regarded as a vortex liquid state due to the presence of unpaired
spinon-vortex composites.\cite{PS1} These spinon-vortices contribute to the
Nernst effect and therefore the lower pseudogap phase should coincide with
the Nernst region discovered \cite{nernst} experimentally in the cuprates.
Recently the electromagnetic response of such a vortex liquid phase has been
also discussed \cite{vortex liquid} based on a different RVB approach.

Previously the magnetic properties in this phase has been discussed \cite%
{PSk} in the context of exploring the driving mechanism in comparison with
the pseudogap phase in the f-RVB state. In the following, we focus on the
magnetic behavior of this lower pseudogap phase and make contrast with that
of the UPP discussed above.

\begin{figure}[tbp]
\begin{center}
\includegraphics[width=3.5in]
{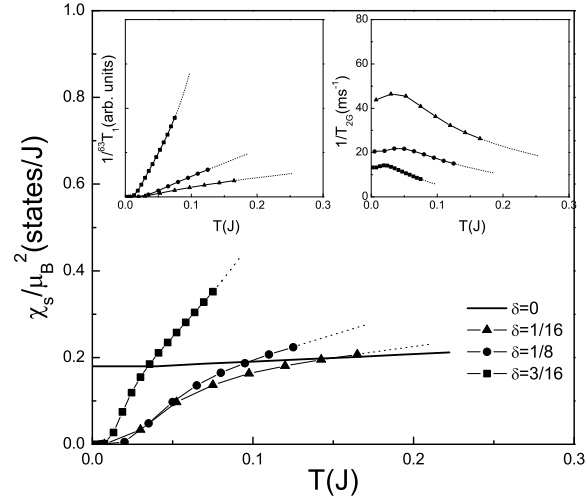}
\end{center}
\caption{Uniform spin susceptibility in the \emph{lower} pseudogap phase at
different dopings including half-filling. The left insert shows $%
1/^{63}T_{1} $ and the right insert $1/T_{2G}$ with the same symbols as in
the main panel. }
\label{6}
\end{figure}

The mean-field equations are the same as in the UPP, but due to the holon
condensation the gauge field $A_{ij}^{h}$ in Eq. (\ref{eig1}) can be treated
as a uniform flux of strength $\pi \delta $ per plaquette \cite{PSk} as
discussed in Sec. II A. In the main panel of Fig. \ref{6}, the uniform spin
susceptibility shows a true \textquotedblleft spin gap\textquotedblright\
behavior, in contrast to the \textquotedblleft scaling\textquotedblright\
curve shown in the UPP in Fig. 4 where $\chi _{s}$ in the doped regime
roughly behaves like that at half-filling---in the latter case $\chi _{s}$
saturates to a constant at $T=0$. In the lower pseudogap phase, these $\chi
_{s}$'s can drop below that at $\delta =0$ and vanish at\ $T=0$ as shown in
Fig. \ref{6}. Such a lower pseudogap behavior has been indeed observed
experimentally.\cite{ps1Bi2212,p.a.li}

Furthermore, in this lower pseudogap phase, $1/^{63}T_{1}$ also decreases
with temperature (see the left inset of Fig. 7), as opposed to the behavior
in the UPP, indicating the appearance of the spin gap over whole momenta. On
the other hand, although the low-energy spin fluctuations are gapped, the
static AF spin-spin correlations as described by the real part of spin
susceptibility function still remain, as reflected by $1/T_{2G}$ shown in
the right inset of Fig. 7, where the monotonic increase of $1/T_{2G}$ in the
UPP (Fig. 6) is replaced by the saturation at lower pseudogap phase. This
feature has also been observed experimentally.\cite{T2G2,T2G3,T2G4}

\section{conclusion and discussion}

In this paper, we have systematically analyzed the magnetic
characterizations of a high-temperature \emph{intrinsic} phase of the
bosonic RVB state, which is described by the formation of the bosonic RVB
order parameter at a temperature below the characteristic $T_{0},$ but still
higher than those for low-temperature orders, including AFLRO and
superconductivity, to emerge.

Such a phase exhibits the pseudogap features that match those of the upper
pseudogap phase in the high-$T_{c}$ cuprate superconductors very well. The
key feature in the crossover to the UPP from above $T_{0}$ is the onset of
the development of strong AF spin-spin correlations, which remain rather
weak at $T>T_{0}$ where the system resembles more an ensemble of
uncorrelated localized spins. This explains why experimentally the uniform
spin susceptibility shows an approximately Curie-Weiss behavior at $T>T_{0},$
reaches a peak at $T_{0}$, and then gets reduces below $T_{0}$ as the weight
of the spin-spin correlations at the momentum $(0,0)$ being transferred to $%
(\pi ,\pi )$, in contrast to an equal weight distribution above $T_{0}$. It
further explains why the spin-lattice relaxation rate gets enhanced for the
planar copper nuclear spins whereas reduced for the planar oxygen nuclear
spins below $T_{0},$ and why the spin-echo decay rate increases with the
decreasing temperature; Clearly the development of the AF correlations is
the underlying mechanism here.

We emphasize that the formation of spin singlet pairing and the onset of AF
correlations at the same time are not always true. In the f-RVB description
of the slave-boson approach, the formation of the f-RVB order parameter
actually leads to the reduction of the spin-lattice relaxation rates for
both the copper and oxygen nuclear spins at low temperatures. This is
because the spin pseudogap opens for both the ferromagnetic and AF
correlations. By contrast, this case occurs in the b-RVB description only at
a lower temperature when the system enters the lower pseudogap regime,
characterized by the formation of Cooper pair amplitude in the so-called
spontaneous vortex phase \cite{PS1} which is a vortex liquid state, short of
superconducting phase coherence. A comparative study of this lower pseudogap
phase in the b-RVB theory and the pseudogap phase in the f-RVB theory has
been given in Ref. \cite{PSk} where two opposite driving mechanisms: kinetic
vs. superexchange energy driven, have been identified.

The phase diagram of the UPP has been determined by a generalized
\textquotedblleft mean-field\textquotedblright\ description in the b-RVB
theory in the parameter space of temperature, doping, and magnetic field.
Since the UPP essentially reflects the spin singlet pairing, the Zeeman
splitting competes directly, which results in a quantitative prediction for
experiment as discussed in Sec. II B.

A central consequence of an RVB (spin singlet) description of the pseudogap
phase is that the pseudogap represents the suppression of the low-lying
spectral weight in spin excitations, but not in charge excitations, as
pointed out in the Introduction. In the following, we briefly outline the
scenario about what happens to the charge channel when one enters the UPP in
the bosonic RVB theory. We shall leave the more quantitative investigation
in future work.

In order to see how the charge transport is affected by the spin
fluctuations, we first note that in the b-RVB theory, the charge degrees of
freedom (holons) are described by the following effective Hamiltonian:\cite%
{PSMFd}%
\begin{equation}
H_{h}=-t_{h}\sum_{<ij>\sigma }e^{iA_{ij}^{s}}h_{i}^{\dagger }h_{j}+h.c.
\label{hh}
\end{equation}%
where the bosonic holons, created by $h_{i}^{\dagger }$, interact with the
gauge field $A_{ij}^{s}$ associated with the spin degrees of freedom.
Similar to the definition of the gauge field $A_{ij}^{h}$ in Eq.(\ref{ah}), $%
A_{ij}^{s}$, which is introduced in the phase string representation (see
Appendix A), satisfies

\begin{equation}
\sum_{<ij>\in C}A_{ij}^{s}=\pi \sum_{l\in \Sigma _{C}}\left( n_{l\uparrow
}^{b}-n_{l\downarrow }^{b}\right) \text{ \ \ }  \label{as}
\end{equation}%
where $C$ is an arbitrary path. Physically $A_{ij}^{s}$ describes $\pm \pi $
flux tubes bound to $\uparrow $($\downarrow $) spinons as seen by holons. So
at $T>T_{0}$, uncorrelated localized spins imply a maximum scattering to the
holons according to Eqs.(\ref{hh}) and (\ref{as}). By forming the RVB
pairing below $T_{0}$, one can easily understand that the fluctuations in $%
A_{ij}^{s}$ will be effectively reduced, and so does the scattering to the
holons according to Eqs.(\ref{hh}) and (\ref{as}), leading to a pseudogap
feature in the charge transport without involving a charge gap. Note that
when the temperature is further reduced to $T_{v}$, where the holons gain
the phase coherence and become Bose condensed, the system enters the lower
pseudogap phase (spontaneous vortex phase) in which the effect of the holon
condensation will feedback to the spinon part via $A_{ij}^{h}\,$in Eq.(\ref%
{hs}) and cause the lower pseudogap phenomenon in the spin part as discussed
in Sec. II E. Finally, the quasiparticle excitation can be regarded as a
recombination of holon, spinon, and phase string, in the superconducting
phase.\cite{qp} It has been argued that the deconfinement occurs above $%
T_{c} $, and the composite structure is expected to shows up in both the
lower and upper pseudogap phases, and the pseudogap feature is thus believed
to be associated with that in the spinon degrees of freedom.

\begin{acknowledgments}
We thank helpful discussions with V. N. Muthukumar, T. Senthil, D. N. Sheng,
H.H. Wen. The authors also acknowledge the support from the grants of NSFC
and MOE. This research was also supported in part by the NSF under Grant No.
Phy99-07949 via KITP at UC Santa Barbara where this work was completed.
\end{acknowledgments}

\appendix

\section{Phase String Formulation}

The $t-J$ model
\begin{equation}
H_{t-J}=-t\sum_{<ij>\sigma }(c_{i\sigma }^{\dagger }c_{j\sigma
}+h.c.)+J\sum_{<ij>}(\vec{S}_{i}\cdot \vec{S}_{j}-\frac{1}{4}n_{i}n_{j})
\label{tJ}
\end{equation}%
may be reformulated by using the phase string decomposition\cite{PS}
\begin{equation}
c_{i\sigma }=(-\sigma )^{i}h_{i}^{\dagger }b_{i\sigma }e^{i\Theta _{i\sigma
}^{string}}  \label{cop}
\end{equation}%
where $h_{i}$ and $b_{i\sigma }$ are all \emph{bosonic fields}. Here $\Theta
_{i\sigma }^{string}$ is a non-local phase factor to restore the fermionic
statistics of the electron operator, and can be expressed as $\Theta
_{i\sigma }^{string}\equiv \frac{1}{2}[\Phi _{i}^{b}-\sigma \Phi _{i}^{h}]$
with $\Phi _{i}^{b}=\sum_{l\neq i}\theta _{i}(l)\left( \sum_{\alpha }\alpha
n_{l\alpha }^{b}-1\right) $, $\Phi _{i}^{h}=\sum_{l\neq i}\theta
_{i}(l)n_{l}^{h}$. Here $\theta _{i}(l)$ is defined as an angle $\theta
_{i}(l)=$ Im $\ln (z_{i}-z_{l})$ with $z_{i}=x_{i}+iy_{i}$ representing the
complex coordinate of a lattice site i. The resulting Hamiltonian $%
H_{t-J}=H_{t}+H_{J}$ reads
\begin{eqnarray}
H_{t} &=&-t\sum_{<ij>\sigma }(e^{i(A_{ij}^{s}-\phi
_{ij}^{0})})h_{i}^{\dagger }h_{j}(e^{i\sigma A_{ij}^{h}})b_{j\sigma
}^{\dagger }b_{i\sigma }+h.c.  \nonumber \\
H_{J} &=&-\frac{J}{2}\sum_{<ij>\sigma \sigma ^{\prime }}(e^{i\sigma
A_{ij}^{h}})b_{i\sigma }^{\dagger }b_{j-\sigma }^{\dagger }(e^{i\sigma
^{\prime }A_{ji}^{h}})b_{j-\sigma ^{\prime }}b_{i\sigma ^{\prime }}
\label{ps}
\end{eqnarray}%
under the no-double-occupancy constraint
\begin{equation}
h_{i}^{\dagger }h_{i}+\sum_{\sigma }b_{i\sigma }^{\dagger }b_{i\sigma }=1%
\text{ \ \ .}  \label{constraint}
\end{equation}%
In the new Hamiltonians, $\phi _{ij}^{0}$ is a $\pi $ flux link variable,
while $A_{ij}^{s}$ and $A_{ij}^{h}$ are constrained by the following
conditions:
\begin{eqnarray}
\sum_{C}A_{ij}^{s} &=&\pi \sum_{l\in C}(n_{l\uparrow }^{b}-n_{l\downarrow
}^{b})  \label{cond1} \\
\sum_{C}A_{ij}^{h} &=&\pi \sum_{l\in C}n_{l}^{h}  \label{cond2}
\end{eqnarray}%
where $C$ is an arbitrary counterclockwise closed path. Finally, in the
phase string representation, the spin operators can be easily reexpressed
according to the decomposition (\ref{cop}) as
\begin{eqnarray}
S_{i}^{z} &=&\frac{1}{2}\sum_{\sigma }\sigma b_{i\sigma }^{\dagger
}b_{i\sigma }\text{ \ \ ,}  \nonumber \\
S_{i}^{\sigma } &=&(-1)^{i}b_{i\sigma }^{\dagger }b_{i-\sigma }e^{i\sigma
\Phi _{i}^{h}}\text{ \ \ .}  \label{sop}
\end{eqnarray}

\section{$1/T_{1}$ Formulation in the Bosonic RVB representation}

$1/T_{1}$ defined in (\ref{NMRG1}) can be reexpressed in terms of the
real-space spin correlation function as follows
\begin{equation}
\frac{1}{T_{1}}=\frac{2k_{B}T}{g^{2}\mu _{B}^{2}N}\sum_{ij}M_{ij}\left.
\frac{\chi _{zz}^{\prime \prime }(i,j,\omega )}{\omega }\right\vert _{\omega
\rightarrow 0^{+}}
\end{equation}%
where $M_{ij}$ is the Fourier transformation of $F(\mathbf{q})^{2}$ in real
space:
\begin{equation}
^{63}M_{ij}\equiv (A_{\perp }^{2}+4B^{2})\delta _{i,j}+2A_{\perp }B\sum_{%
\hat{\eta}}\delta _{i,j+\hat{\eta}}+B^{2}\sum_{\hat{\eta}\neq -\hat{\eta}%
^{\prime }}\delta _{i,j+\hat{\eta}+\hat{\eta}^{\prime }}
\end{equation}%
where $\hat{\eta}=\pm \hat{x},\pm \hat{y}$ and
\begin{equation}
^{17}M_{ij}\equiv 2C^{2}(\delta _{i,j}+\frac{1}{4}\sum_{\hat{\eta}}\delta
_{i,j+\hat{\eta}})\text{ \ \ \ .}
\end{equation}%
From $M_{ij}$ we see that only up to the next-nearest-neighbor spin
correlations are involved in the spin-lattice relaxation rates of $^{63}Cu$
and $^{17}O$ nuclear spins.

In the Bosonic RVB mean-filed theory, by using Eq.(\ref{bogoliubov}) the
dynamic spin susceptibility can be expressed as
\begin{equation}
\left. \frac{\chi _{zz}^{\prime \prime }(i,j,\omega )}{\omega }\right\vert
_{\omega \rightarrow 0^{+}}=G_{ij}^{-}+(-1)^{i-j}G_{ij}^{+}  \label{dsus1}
\end{equation}%
where
\begin{equation}
G_{ij}^{\pm }=\frac{\pi }{2}{\sum_{mm^{\prime }}}^{\prime }K_{mm^{\prime
}}^{zz}(i,j)\left( -\frac{\partial n(E_{m})}{\partial E_{m}}\right)
(p_{mm^{\prime }}^{\pm })^{2}\delta (E_{m}-E_{m^{\prime }})  \label{dsus2}
\end{equation}%
in which ${\sum }^{\prime }$ denote the summation of $m$ with $\xi _{m}>0$
and
\begin{equation}
K_{mm\prime }^{zz}(i,j)\equiv \sum_{\sigma }w_{m\sigma }^{\ast
}(i)w_{m\sigma }(j)w_{m^{\prime }\sigma }^{\ast }(j)w_{m^{\prime }\sigma }(i)
\label{dsus3}
\end{equation}

By noting that $p_{mm^{\prime }}^{-}=1,p_{mm^{\prime }}^{+}=\lambda /E_{m}$
at $E_{m}=E_{m^{\prime }}$, we can further reexpress $1/T_{1}$ in the
following form
\begin{equation}
\frac{1}{T_{1}}=\frac{2}{3g^{2}\mu _{B}^{2}N}{\sum_{m}}^{\prime
}n(E_{m})(1+n(E_{m}))\rho (E_{m})\left[ D_{m}^{-}+\frac{\lambda ^{2}}{%
E_{m}^{2}}D_{m}^{+}\right]  \label{dsus4}
\end{equation}%
where the density of states $\rho (E_{m})=(2/N){\sum_{m}}^{\prime }\delta
(E_{m}-E_{m^{\prime }})$ and the coefficient, $D_{m}^{\pm }$, is defined by
\begin{equation}
D_{m}^{\pm }=\frac{{\sum_{m}}^{\prime }d_{mm^{\prime }}^{\pm }\delta
(E_{m}-E_{m^{\prime }})}{{\sum_{m}}^{\prime }\delta (E_{m}-E_{m^{\prime }})}
\label{dsus5}
\end{equation}%
with
\begin{equation}
d_{mm^{\prime }}^{\pm }\equiv \frac{\pi }{2}N\sum_{ij}K_{mm^{\prime
}}^{zz}(i,j)(\mp )^{i-j}M_{ij}  \label{dsus6}
\end{equation}

In Eq.(\ref{dsus4}) a numerical factor $2/3$ is also added just like the
uniform spin susceptibility as noted in the main text as at half-filling.%
\cite{half} The final result will be an average over different random
configurations of $A_{ij}^{h}$ due to the incoherent distribution of holes
in the UPP (In the lower pseudogap phase, by contrast, the holon
condensation leads to a uniform flux distribution of $A_{ij}^{h}$ and no
such average is needed$).$ The calculation is done on a $32\times 32$
lattice and results are presented in Fig. \ref{fnmr}.\cite{remark1}

\section{spin-echo relaxation rate}

The spin-echo relaxation rate $1/T_{2G}$ is defined in Eq.(\ref{T2}). In the
b-RVB theory, the real part of the static susceptibility, $\chi
_{zz}^{\prime }(\mathbf{q})$, can be expressed as
\begin{equation}
\chi _{zz}^{\prime }(\mathbf{q})=\chi _{zz}^{\prime +}(\mathbf{q})+\chi
_{zz}^{\prime -}(\mathbf{q})  \label{susq1}
\end{equation}%
with
\begin{widetext}
\begin{equation}
\chi^{\prime\pm}_{zz}(\textbf{q})=\frac{2}{3}\times\frac{1}{2}{\sum_{mm^\prime}}^\prime
K_{mm^\prime}^{zz}(\textbf{q})\left[(p_{mm^\prime}^\pm)^2\frac{(n(E_{m^\prime})-n(E_m))}
{E_m-E_{m^\prime}}+(l_{mm^\prime}^\pm)^2\frac{(1+n(E_m)+n(E_{m^\prime}))}
{E_m+E_{m^\prime}}\right]\label{susq2}
\end{equation}
\end{widetext}where
\begin{eqnarray}
p_{mm^{\prime }}^{\pm } &=&u_{m}u_{m^{\prime }}\pm v_{m}v_{m^{\prime }}
\nonumber \\
l_{mm^{\prime }}^{\pm } &=&u_{m}v_{m^{\prime }}\pm v_{m}u_{m^{\prime }}
\label{susq3}
\end{eqnarray}%
and
\begin{equation}
K_{mm^{\prime }}^{zz}(\mathbf{q})\equiv \frac{1}{N}\sum_{ij\sigma }e^{i%
\mathbf{q}\cdot (\mathbf{r}_{i}-\mathbf{r}_{j})}w_{m\sigma }^{\ast
}(i)w_{m\sigma }(j)w_{m^{\prime }\sigma }^{\ast }(j)w_{m^{\prime }\sigma }(i)
\label{susq5}
\end{equation}%
The numerical calculation is similar to that for $1/T_{1}$.


\begin{thebibliography}{99}
\bibitem{remark1} Note that in the inset of Fig. 5, $1/^{63}T_{1}T$ is shown
only below $T\leq 0.8T_{0},$ as $1/^{63}T_{1}T$ is ill-defined at $T_{0}$ in
the mean-field description where $\Delta ^{s}=0$. Such an ill-behavior is
also reflected in the uniform susceptibility in Fig. 2 by a sharp cusp at $%
T_{0}$ as compared to the broad, smooth peak in the HTSE result and
experimental data. This is due to the fact that there is no true phase
transition at $T_{0}$ and the description beyond the mean-field order
parameter $\Delta ^{s}$ is needed in order to more accurately describe the
crossover behavior in the transition regime. On the other hand, the ratio $%
^{17}T_{1}/^{63}T_{1}$ still remains well-defined around $T_{0}$ in the
mean-field state as shown in the main panel of Fig. 5.
\end{thebibliography}
\end{document}